\documentclass[12pt,english,american]{article}    
\usepackage[unicode=true,pdfusetitle,
 bookmarks=true,bookmarksnumbered=false,bookmarksopen=false,
 breaklinks=true,pdfborder={0 0 0},backref=false,colorlinks=true]{hyperref}
\hypersetup{linkcolor=blue,citecolor=blue, urlcolor=blue}

\usepackage{color}
\usepackage{amsmath} 
\usepackage{amssymb}
\usepackage{latexsym}
\usepackage{enumitem}
\numberwithin{equation}{section}

\newcommand{\df}{\doteq}
\newcommand{\gc}{\color{black}}

\newcommand{\mrm}{\mathrm}
\newcommand{\1}{\!\!\!}

\newcommand{\out}{\mathrm{out}}

\newcommand{\mcA}{\mathcal A}

\newcommand{\mcF}{\mathcal{F}}
\renewcommand{\Im}{\mathrm{Im}}

\newcommand{\slim}{\te{s-}\lim}

\newcommand{\mcL}{\mathcal{L}}

\newcommand{\lan}{\langle}
\newcommand{\ran}{\rangle}

\newcommand{\supp}{\mathrm{supp}}

\newcommand{\Om}{\Omega}

\newcommand{\h}{\fr{1}{2}}

\newcommand{\te}{\textrm}
\newcommand{\vp}{\varphi}

\newcommand{\de}{\delta}

\newcommand{\hil}{\mathcal{H}}

\newcommand{\mfa}{\mathcal{F}}
\newcommand{\mco}{\mathcal{O}}

\newcommand{\eps}{\varepsilon}
\newcommand{\fr}[2]{\frac{#1}{#2}}
\newcommand{\al}{\alpha}
\newcommand{\be}{\beta}

\newcommand{\real}{\mathbb{R}}

\newcommand{\la}{\lambda}
\newcommand{\ov}{\overline}

\newcommand{\non}{\nonumber}

\def\proof{\noindent{\bf Proof. }}
\def\qed{$\Box$\medskip}

\newtheorem{theoreme}{Theorem } [section]
\newtheorem{proposition}[theoreme]{Proposition}
\newtheorem{lemma}[theoreme]{Lemma}
\newtheorem{definition}[theoreme]{Definition}
\newtheorem{corollary}[theoreme]{Corollary}
\newtheorem{remark}[theoreme]{Remark}
\newtheorem{example}[theoreme]{Example}
\newtheorem{criterion}[theoreme]{Criterion}
\newtheorem{assumption}[theoreme]{Assumption}
\newcommand{\beex}{\begin{example}}
\newcommand{\eeex}{\end{example}}
\newcommand{\bea}{\begin{assumption}}
\newcommand{\eea}{\end{assumption}}
\newcommand{\beqa}{\begin{eqnarray}}
\newcommand{\eeqa}{\end{eqnarray}}
\newcommand{\ben}{\begin{arabicenumerate}}
\newcommand{\een}{\end{arabicenumerate}}
\newcommand{\bex}{\begin{example}}
\newcommand{\eex}{\end{example}}
\newcommand{\ber}{\begin{remark}}
\newcommand{\eer}{\end{remark}}
\newcommand{\bec}{\begin{corollary}}
\newcommand{\eec}{\end{corollary}}
\newcommand{\bed}{\begin{definition}}
\newcommand{\eed}{\end{definition}}
\newcommand{\bep}{\begin{proposition}}
\newcommand{\eep}{\end{proposition}}
\newcommand{\becr}{\begin{criterion}}
\newcommand{\eecr}{\end{criterion}}
\def\bel{\begin{lemma}}
\def\eel{\end{lemma}}
\def\bet{\begin{theoreme}}
\def\eet{\end{theoreme}}
\def\bed{\begin{definition}}
\def\eed{\end{definition}}

\newcommand{\rr}{r}

\def\undertilde#1{\mathord{\vtop{\ialign{##\crcr
$\hfil\displaystyle{#1}\hfil$\crcr\noalign{\kern1.5pt\nointerlineskip}
$\hfil\tilde{}\hfil$\crcr\noalign{\kern1.5pt}}}}}
\def\wideundertilde#1{\mathord{\vtop{\ialign{##\crcr
$\hfil\displaystyle{#1}\hfil$\crcr\noalign{\kern1.5pt\nointerlineskip}
$\hfil\widetilde{}\hfil$\crcr\noalign{\kern1.5pt}}}}}

\newcommand{\RR}{\mathbb{R}}

{\par \medskip \noindent {\em Proof.}}{\hspace*{\fill} $\square$\par%
\medskip\noindent}
{\par \medskip \noindent {\em Proof of\, }}{\hfill $\square$\par%
\medskip\noindent}

\newcommand{\bsp}{p^1{}}

\renewcommand{\Im}{\text{Im }}

\newcommand{\calF}{\mathcal{F}}

\newcommand{\calD}{{\mathcal D}}

\newcommand{\calP}{{\mathcal P}}

\newcommand{\unity}{{\setlength{\unitlength}{1em}
                     \begin{picture}(0.75,1)
                     \put(0,0){$1$}
                     \put(0.34,0){\line(0,1){0.65}}
                     \end{picture}
                   }}

\newcommand{\Wickli}{:\!}
\newcommand{\Wickre}{\!:}
\newcommand{\TwoPt}{w}
\newcommand{\Comm}{iD_0}  

\newcommand{\reg}{v}  

\newcommand{\bos}{{}}  

\newcommand{\WickReg}[1]{\Wickli W(#1)\Wickre_\reg}

\newcommand{\beq}{\begin{equation}}
\newcommand{\eeq}{\end{equation}}
\newcommand{\sgn}{\text{\rm sgn}}

\begin{document} 
\title{Interacting massless infraparticles 
    in 1+1 dimensions} 
\author{Wojciech\ Dybalski$^1$ and  Jens\ Mund$^2$ \\\\
$^1$ Faculty of Mathematics and Computer Science \\  Adam Mickiewicz University in Pozna\'n\\
ul. Uniwersytetu Pozna\'nskiego 4, 61-614 Pozna\'n, Poland\\
\small{E-mail: {\tt wojciech.dybalski@amu.edu.pl}} \\\\
$^2$ Dept.\ de F\'{\i}sica, Universidade Federal de Juiz de Fora, \\
$36036-900$ Juiz de Fora, MG, Brazil. 
 \\
\small{E-mail: {\tt   mund@ufjf.fisica.br }}\\\\  
}
\date{}
\maketitle
\begin{abstract}
 The Buchholz' scattering theory of waves in two dimensional massless models suggests a natural
 definition of a scattering amplitude.   We compute such
 a scattering amplitude for charged infraparticles  that live in the  GNS representation of the 2$d$ massless scalar free field  and obtain a non-trivial result. It turns out that these  excitations exchange phases,
 depending on their charges, when they collide.    
  \end{abstract}

\section{Introduction}

Construction of interacting quantum field theories, even in low-dimensional spacetime, is  notoriously
difficult, cf. e.g.  \cite{GJ, Le08,Ta14, CT15, Di18, GH21, GHW20}. Apart from the well known technical problems  there is also a  conceptual difficulty:
what does it mean that a quantum field theory is interacting?  Non-triviality of the $S$-matrix  is a clear-cut
criterion only for collisions of Wigner particles\footnote{We refer to Chapters II.4 and VI of \cite{Ha} for classical results and to \cite{Dy05, DH15, Du17, Du18} for more recent developments.}.  
Outside of this restrictive setting, excluding most physical particles \cite{Bu86},  there is no generally accepted criterion for interaction and 
one has to proceed on a case by case basis.
In this note we formulate a natural criterion for interaction for massless two-dimensional
theories, which is suggested by the Buchholz' collision theory of waves \cite{Bu75}. It allows us to exhibit a subtle scattering between charged excitations in massless free field theory in two dimensions, which consists in exchanging
phases depending on their charges.  This interaction, which coexists with a linear field equation, is  effected by the exotic infrared structure 
of the vacuum in the model.  We remark that a similar effect was anticipated by Streater in \cite{St10}.

Let us now introduce our criterion for interaction for Haag-Kastler theories $(\mfa, U, \Om)$ of massless particles on $\real^2$.
Here $\mfa\subset B(\hil)$ is a $C^*$-algebra\footnote{It will be interpreted as the $C^*$-algebra of charge carrying fields. The sub-algebra of observables $\mathcal{A}$ will appear in a concrete model  in Section~\ref{Bose}.} acting irreducibly on a Hilbert space $\hil$, $U$ is a strongly continuous unitary representation of the Poincar\'e group satisfying the spectrum
condition  and the unit vector $\Om\in \hil$  is the vacuum vector, that is, $\Om$ is cyclic for $\mfa$ and invariant under $U$. Furthermore, $\mfa$ is the global
algebra of a net of $C^*$-algebras $\mco \mapsto \mfa(\mco)$, labelled by open bounded regions $\mco\subset \real^2$,  which is local and isotonous.  
Following \cite{Bu75}, for any $F\in \mfa$ we define the averaged operators at time $|T|\geq 1$
\beqa
F_{\pm}(h_{T})=\int dt\, h_{T}(t) F(t_{\pm}), \quad t_{\pm}\df(t, \pm t), 
\eeqa
where $h\in \mathcal{D}_{\real}(\real)$, $h\geq 0$, $h(t)=h(-t)$,  $\int dt h(t)=1$ and $h_{T}(t)\df \fr{1}{s(T)} h\big(\fr{t-T}{s(T)}   \big)$,
$s(T)\df\ln|T|$. For any $F, G\in \mfa$ we define vectors
\beqa
\Psi_T\1&\df&\1F_{+}(h_T) G_-(h_T)\Om\in \hil, \label{wave-vectors} \\
\Psi_T^{\otimes} \1&\df&\1 F_{+}(h_T)\Om\otimes G_-(h_T)\Om\in \hil \otimes \hil.
\eeqa 
In this work we are interested in  \textbf{scattering amplitudes} of massless excitations of the net $(\mfa, U, \Om)$  which are defined by
\beqa
S(F,G)\df\lim_{T\to \infty} \fr{\lan \Psi_{T}, \Psi_{-T}\ran }{\lan \Psi^{\otimes}_T,   \Psi^{\otimes}_{-T} \ran } 
=\lim_{T\to \infty} \fr{\lan \Om, G_-^*(h_T)F_{+}^*(h_T) F_{+}(h_{-T}) G_-(h_{-T})\Om\ran } 
{\lan \Om, G_-^*(h_T) G_-(h_{-T}) \Om\ran \lan \Om, F_{+}^*(h_T) F_{+}(h_{-T}) \Om  \ran}, \label{amplitude-def}
\eeqa
provided that the denominator is non-zero for sufficiently large $T$ and the limit exists and satisfies $|S(F,G)|\leq 1$.
If, in addition, $0\neq S(F,G)\neq 1$
for some $F,G\in \mfa$, then we say that the Haag-Kastler theory is \textbf{interacting}.   The denominator in (\ref{amplitude-def}) ensures that $S(F,\unity)=S(\unity,G)=1$. Hence, like in the conventional setting discussed below,  there is no one-body interaction.

Let us now point out that also at the two-body level this concept of interaction generalizes the conventional one from \cite{Bu75}.
The latter work  studies vectors~(\ref{wave-vectors}) under the assumption that the invariant subspaces $\hil_{\pm}$
of $t\mapsto U(t_{\pm})$ contain some non-zero vectors orthogonal to the vacuum. These vectors describe massless particles in the sense of Wigner, 
which in this context are called waves. For any non-zero $\Psi_{\pm}\in \hil_{\pm}$
one can find $F, G\in \mfa$ s.t. 
\beqa
\Psi_+=\lim_{T\to \pm\infty} F_{+}(h_T) \Om, \quad  \Psi_-=\lim_{T\to \pm\infty} G_-(h_T)\Om.  \label{single-particle}
\eeqa
Then the limits
\beqa
\Psi^{\mathrm{out}}\df\lim_{T\to \infty} \Psi_{T}=\lim_{T\to \infty}F_{+}(h_T) G_-(h_T)\Om , \quad  
\Psi^{\mathrm{in}}\df\lim_{T\to \infty} \Psi_{-T}= \lim_{T\to \infty}F_{+}(h_{-T}) G_-(h_{-T})\Om \label{scattering-states}
\eeqa
exist, are different from zero, depend only on the single-particle vectors $\Psi_{\pm}$ and satisfy $\,\| \Psi^{\mathrm{out}}\|=\|\Psi_+\otimes \Psi_-\|=\|\Psi^{\mathrm{in}}\|$. Thus one can define the (isometric) scattering matrix by 
\beqa
S\Psi^{\mathrm{out}}= \Psi^{\mathrm{in}}.
\eeqa
Then the scattering amplitude (\ref{amplitude-def}) has the form 
\beqa
S(F,G)=\fr{\lan \Psi^{\out}, S \Psi^{\out}\ran}{\lan \Psi^{\out}, \Psi^{\out}\ran }. \label{conv-amplitude}
\eeqa
In this context $S(F,G)\neq 1$ implies $S\neq \unity$, that is, interaction in the conventional sense. 
We remark as an aside,
that  scattering theory of waves was  generalized to wedge-local Haag-Kastler nets and interacting examples were
found in this broader setting in \cite{DT11, BT13}. 
 
 For $F,G$ s.t.\ the vectors $F\Om$, $G\Om$ are orthogonal to the subspaces  of waves $\hil_{\pm}$, the resulting
 scattering states~(\ref{scattering-states}) vanish. This fact alone should not be  interpreted as interaction but as 
 a manifestation of the infraparticle problem. Our criterion for interaction, $0\neq S(F,G)\neq 1$, accounts for this,
 and may still hold due to cancellations between the numerator and the denominator in (\ref{amplitude-def}).
 Such a  situation occurs for charge carrying fields in massless free field theory in two dimensions and the resulting scattering
 amplitude is computed for specific $F,G$ in   Theorem~\ref{main-result} below. The result is 
 \beqa
 S(F,G)= e^{-\fr{i}{2}q_f q_g},  \label{s-a}
 \eeqa
 where $q_f$, $q_g$  are charges of the excitations created by  $F,G$  from the vacuum. Such excitations can be called
 \emph{infraparticles} as they are not eigenvectors of the relativistic mass operator. While it is well known that massless two dimensional theories contain infraparticles \cite{Sch63,  Bu96, DT12, DT13},  we give the first example in which they interact by exchanging charge-dependent phases in collisions.   Similarly  to the Ising or Federbush model (cf. \cite{Le05, Ta14}),    the scattering amplitude (\ref{s-a})  is independent of momenta.




Our paper is organized as follows: In Section~\ref{Bose} we define the model and state our main result.
Section~\ref{Asymptotics} contains some preparations on the behaviour of vacuum expectation
values under large spacetime translations.  In Section~\ref{last-section} we give the proof of the main result.
 Some technical details of the discussion are postponed to the appendices.
 
\vspace{0.2cm}

\noindent{\bf Acknowledgments.}  Both authors were  partially supported by the  Emmy Noether grant DY107/2-2 of the DFG. W.D.  also acknowledges  support of the NCN within the grant `Sonata Bis' 2019/34/E/ST1/00053. J.M.\ received financial support from the Brazilian research agency CNPq, and is also grateful to CAPES and Finep. 

\section{The model and the main result} \label{Bose}
\setcounter{equation}{0}

We outline the massless scalar free field theory on two-dimensional Minkowski spacetime,  referring to \cite{Abdalla, AMS92, AMS93, BahnsFredRejzner17, BogLogOksTod, SW70, Ci09, Derezinski1+1, Strocchi90, Sch13}  for more information.
We denote the momentum by $\bsp\in\RR$, and the energy-momentum of a particle of mass $m\geq 0$ by $p\df (\omega_m(\bsp),\bsp)$,  $\omega_m(\bsp)\doteq \sqrt{(\bsp)^2+m^2}$. 
The two-point function $\TwoPt_m$ of the free field $\phi_m$ in spacetime dimension $d=2$ with mass $m>0$ is
\begin{equation} \label{eq2Ptm>0}
\TwoPt_m(x) \df (2\pi)^{-1}\int d\mu_m(p) e^{-ip\cdot x},
\end{equation}
where $d\mu_m(\bsp) \doteq d\bsp/2\omega_m(\bsp)$.
The limit $\TwoPt\doteq \lim_{m\to 0}\TwoPt_m$ is only well-defined for test functions from $\mathcal{D}(\RR^{2})$  
whose \textbf{charge} $q_f\df\int f(x)d^2x$ is equal to zero. On the other hand, the commutator function
\begin{equation} \label{eqComm}
\Comm(x) \doteq \lim_{m\to 0} \big(\TwoPt_m(x)-\TwoPt_m(-x)\big)
\end{equation}
is well-defined for all test functions from $\mathcal{D}(\RR^{2})$.   This gives a non-degenerate symplectic form on the vector space
$\mcL\df\mathcal{D}_{\real}(\RR^{2}) / \Box \mathcal{D}_{\real}(\RR^{2})$, where $\Box$ is the d'Alembertian. The non-degeneracy can be shown
 as in \cite[Theorem~3.4.7]{BGP07}    using explicit formulas for the propagators, see e.g.  \cite[formula~(3.3)]{CRV21}.
Thus, by \cite[Theorem 4.2.9]{BGP07}  we obtain  a unique $C^*$-algebra  $\calF_{\bos}$ 
generated by the abstract Weyl operators $W(f)$, $f\in \mathcal{D}_{\real}(\RR^{2})$,   with the Weyl relations
\beqa
 W(f)W(g) = e^{-\Comm(f,g)/2}W(f+g)\quad \textrm{and}\quad  W(f)^*=W(-f),
 \eeqa
 where $D_0(f,g)\doteq \int f(x)D_0(x-y)g(y) d^2xd^2y$.  
 For any open bounded region $\mco\subset \real^2$, the subspace $\mcL(\mco)\df \mathcal{D}_{\real}(\mco) / \Box \mathcal{D}_{\real}(\RR^{2})$  of $\mcL$ gives rise to a $C^*$-algebra $\mcF(\mco)$ which is naturally a subalgebra of $\mcF$. The resulting
 isotonous net $\mco\mapsto \mcF(\mco)$ is  local, since $D_0$ is supported on lightlike vectors.   Denoting by $\calP_+^\uparrow$
 the proper ortochronous  Poincar\'e group, for any  $L\in\calP_+^\uparrow$ we define an automorphism of $\mcF$ by
 \beqa
 \al_L(W(f))\df W(f_L), \quad f_L(x)\df f(L^{-1} x),  \label{Poincare-transformations}
 \eeqa
 which acts covariantly on the net.
Next, we introduce a linear functional  $\langle\; \cdot \;\rangle$ on $\calF_\bos$ by
\begin{equation} \label{eqVac}
 \langle W(f) \rangle \doteq \lim_{m\to 0}e^{-\frac{1}{2}\TwoPt_m(f,f)}=\begin{cases} e^{-\frac{1}{2}\TwoPt(f,f)}&\text{ if } q_f =0, \quad \\ 0 & \text{ else. }
 \end{cases}
\end{equation}
The above definition is consistent with the fact that  $\TwoPt_m(f,f)\df \int f(x)w_m(x-y)f(y)d^2xd^2y$  diverges to $+\infty$ with $m\to 0$ if $q_f\neq 0$. By \cite{AMS92, AMS93} this functional defines a state on $\mcF$ and the resulting GNS representation 
$(\pi, \hil, \Om)$
 is irreducible. This state is invariant under the Poincar\'e transformations~(\ref{Poincare-transformations}) and the corresponding group of automorphisms  is unitarily implemented by a strongly continuous group of unitaries $U$  on $\hil$ satisfying the spectrum condition (cf. Appendix~\ref{GNS}). Thus  the functional (\ref{eqVac}) is a pure vacuum state on $\mcF$ in the sense of \cite{Ar} and $(\pi(\mcF), U,\Om)$ is a Haag-Kastler net as defined in the Introduction. 

Let $\mcA\subset \mcF$ be the subalgebra generated by $W(f)$, $q_f=0$.  We can use it to decompose the Hilbert 
space into the neutral and charged subspace:
\beqa
\hil=\hil_0\oplus \hil_{\mrm{ch}}, \quad \hil_0 \df \ov{\pi(\mcA)\Om}, \quad \hil_{\mrm{ch}}\df \hil_0^{\bot}.
\eeqa 
By cyclicity of $\Om$, the subspace $\hil_{\mrm{ch}}$ is spanned by all $W(f)$ s.t. $q_f\neq 0$. As shown in  Appendix~\ref{GNS}
\beqa
\hil_{\pm} \subset \hil_0 \label{waves-are-neutral}
\eeqa
thus all the single-wave vectors  in the model have zero charge.
In Section~\ref{last-section} of this paper we compute the scattering amplitude (\ref{amplitude-def}) both for neutral and charged excitations in this theory and obtain the following result:
\bet\label{main-result} Let $(\pi(\mfa), U, \Om)$ be the Haag-Kastler net of the massless scalar free field, as defined above. 
Then, for $F=\pi(W(f))$, $G=\pi(W(g))$, $f,g\in \mathcal{D}_{\real}(\real^2)$,  the scattering amplitude (\ref{amplitude-def}) exists and equals
\beqa
S(F, G)=e^{-\fr{i}{2}q_f q_g}, \label{amplitude}
\eeqa 
where $q_f\df \int f(x)d^2x $ and $q_g\df \int g(x) d^2x$.
\eet

We infer from this theorem that colliding excitations in the model exchange phases depending on their charges.
Clearly, a non-trivial effect occurs only if both excitations are charged, i.e. $q_fq_g\neq 0$. In this case neither
of them can be a wave, cf. (\ref{waves-are-neutral}),  thus they should be considered infraparticles.  To resolve the apparent
paradox that a free field is interacting, one should recall that the $2d$ massless free field has a much reacher vacuum structure
than its higher dimensional counterparts: The representation $\pi$ is not regular, so the field $\phi$ does not exist. The Hilbert
space $\hil$ is not separable, so it is not the usual Fock space. In fact it is an uncountable direct sum of Fock spaces labelled
by the charges, on which the  operators $\pi(W(f))$ act in a way dictated by  $q_f$ (cf.  \cite[formula (4.4)]{AMS93}).   
Considering all this, it is not a surprise that `free field' turns out to be a misnomer for this model.


\section{Spacetime  asymptotics of the vacuum state} \label{Asymptotics}
\setcounter{equation}{0}

A standard tool to deal with the state (\ref{eqVac}) is  the regularized two-point function: 
\begin{equation} \label{eq2PtReg'}
  \TwoPt_{\reg}(x)\doteq (2\pi)^{-1}\int d\mu_0(p) \big(e^{-ip\cdot x} - v(p)\big),
\end{equation}
where $v: \real^2\to \real$ is a measurable function s.t.  $v(|p^1|,p^1)=v(|p^1|,-p^1)$,
$v(0,0)=1$ and for some $r,\epsilon, c>0$
\beqa
|v(|\bsp|,\bsp)-v(0,0)|\leq c|\bsp|^{\epsilon} \textrm{ for } |\bsp|<r
\textrm{ and }\int_{|\bsp|>r} \fr{d\bsp}{|\bsp|}  | v(|\bsp|,\bsp)|<\infty. \label{technical}
\eeqa
(Note that different functions $v$ give rise to an additive constant, $w_{v'}(x)=w_v(x)+c'$).  
Due to (\ref{technical}), $w_v(f,f)$ is finite for all $f\in \mathcal{D}_{\real}(\real^2)$ and we can define the regularized Wick exponentials  by
\begin{equation} \label{eqVacReg}
\WickReg{f} \; \doteq \;  \frac{W(f)}{\langle W(f)\rangle_\reg}, \textrm{ where } \langle W(f)\rangle_\reg \doteq e^{-\frac{1}{2}\TwoPt_\reg(f,f)}. 
\end{equation}
Needless to say, they generate the same $C^*$-algebra $\mcF$. The restriction of the regularized functional $\langle\;\cdot\; \rangle_\reg$ to the neutral subalgebra $\mcA$ coincides with the physical vacuum state $\langle \;\cdot \; \rangle$, since  
$\TwoPt(f,f)=\TwoPt_\reg(f,f)$ if $q_f=0$, cf. (\ref{eq2PtReg'}).
However, the regularized functional is non-positive \cite[p.27]{Abdalla} and should only be seen as a device for organizing computations.
In particular, it is easy to see that the Weyl relations and the vacuum expectation values of products of regularized Wick exponentials
have the form 
\begin{align} \label{eqWickThmReg}
  \WickReg{f_1}\cdots \WickReg{f_n} \; & = e^{-\sum_{i<j} \TwoPt_\reg(f_i,f_j)}\, \WickReg{\sum_{i=1}^n f_i}, \quad  \\
\langle \WickReg{f_1}\cdots \WickReg{f_n}\rangle  \; & = \delta_{0,q} \; e^{-\sum_{i<j} \TwoPt_\reg(f_i,f_j)},\quad q\doteq \sum_{i=1}^{n} q_{f_i}. \label{eqWickThmRegVEV}
\end{align}
Equality  (\ref{eqWickThmRegVEV}) follows from $\langle \WickReg{f}\rangle = \delta_{0,q_f}$ and we stress
that there is the physical (non-regularized) vacuum state (\ref{eqVac}) on its left hand side.

It is well known \cite[p.27]{Abdalla} and very useful for our investigation that (\ref{eq2PtReg'}) can be rewritten in the form
\begin{equation} \label{eq2PtReg}
\TwoPt_{\reg}(x)= { \lim_{\eps\downarrow 0 }}\frac{-1}{4\pi} \ln(-\mu_v^2 x^2 + i \eps x^0),
\end{equation}
where the scale $\mu_v>0$ depends on $v$ and the limit is in $\mathcal{D}'(\real^2)$. The precise definition of this logarithm is often glossed over in the literature. Here and below  the complex logarithm will always be understood with a cut along the negative real axis. As the imaginary part of this logarithm  gives rise to the non-trivial amplitude in (\ref{amplitude}), we provide a detailed proof   of the equality between (\ref{eq2PtReg'}) and (\ref{eq2PtReg})  in Appendix~\ref{2PtBoson}.

Using (\ref{eq2PtReg}) we will study the asymptotic behaviour of the regularized two-point function in  
Lemmas~\ref{log-lemma}, \ref{2Pttv} below. Then, exploiting  formula~(\ref{eqWickThmRegVEV}), we will  compute the scattering amplitude (\ref{amplitude-def}) in  the next section.
\bel\label{log-lemma} Let   $f\in \mathcal{D}(\real^2)$ and $h_1, h_2$ be real polynomials in $x^0,x^1$ which are not identically zero. Then
\beqa
\lim_{\eps\downarrow 0} \int \ln[h_1(x)+i\eps h_2(x)] f(x) d^2x=\int  \big( \ln|h_1(x)|  + \theta(-h_1(x)) \mrm{sgn}(h_2(x)) i\pi     \big)f(x)d^2x.  \label{ln-lemma}
\eeqa
\eel
\proof  For $0<\eta<1$,  denote  $N(\eta) \df \{\, x\in \supp(f)\,|\, |h_1(x)|<\eta\}$ and $N(\eta)'\df \real^2\backslash N(\eta)$. 
We decompose the region of integration:
\beqa
\int \ln[h_1(x)+i\eps h_2(x)] f(x) d^2x\1&=&\1 \int_{N(\eta)} \ln[h_1(x)+i\eps h_2(x)] f(x) d^2x  \label{rest-term} \\
 \1& &\1  + \int_{N(\eta)'} \ln[h_1(x)+i\eps h_2(x)] f(x) d^2x. \label{leading-term}
\eeqa
We intend to first take the limit $\eps\downarrow 0$ and then $\eta\downarrow 0$. In the leading term (\ref{leading-term}) the limit $\eps\downarrow 0$ can
be computed by the dominated convergence, which gives
\beqa
& &\lim_{\eps\downarrow 0}\int_{N(\eta)'} \ln[h_1(x)+i\eps h_2(x)] f(x) d^2x \non\\
& &= \int_{N(\eta)'}  \big(\theta(  h_1(x) ) \ln|h_1(x)|  + \theta(-h_1(x))( \ln|h_1(x)|+\mrm{sgn}(h_2(x)) i\pi )    \big)f(x)d^2x \non\\
& &= \int_{N(\eta)'}  \big( \ln|h_1(x)|  + \theta(-h_1(x)) \mrm{sgn}(h_2(x)) i\pi     \big)f(x)d^2x,
\eeqa
where $\theta$ is the Heaviside function.
Now in the limit $\eta\downarrow 0$ dominated convergence gives the expression on the r.h.s. of (\ref{ln-lemma}). 

Now we consider the rest term~(\ref{rest-term}).  We rewrite it as follows
\beqa 
\int_{N(\eta)} \ln[h_1(x)+i\eps h_2(x)] f(x) d^2x\1& =&\1 \int_{N(\eta)} \h \ln[ h_1(x)^2+\eps^2 h_2(x)^2 ]f(x) d^2x \label{rest-term-one}\\
\1&  &\1+ \int_{N(\eta)}  i\vp_{\eps}(x)   f(x) d^2x, \label{rest-term-two}
\eeqa
where $\vp_{\eps}(x)$ is the phase of  $h_1(x)+i\eps h_2(x)$. By dominated convergence, expression~(\ref{rest-term-two})
has the limit $\eps\downarrow 0$ equal to $ \int_{N(\eta)} (\pm i\pi) f(x) d^2x$, where $\pm$ may depend on $x$.  The subsequent 
limit $\eta\downarrow 0$ is equal to zero due to the shrinking of the region of integration. As for (\ref{rest-term-one}), after decomposing
the region of integration into $N_{\pm}(\eta) \df  \{x\in N(\eta) \,|\, \pm f(x)\geq 0\,  \}$ and choosing $\eps$ small enough,
so that $h_1(x)^2+\eps^2 h_2(x)^2\leq 1$ in the region of integration (recall that $\eta<1$) we  use the monotone convergence
to compute the limit $\eps\downarrow 0$. The resulting expression $\int_{N(\eta)} \h \ln[h_1(x)^2]f(x) d^2x $ tends also to zero
with $\eta \downarrow 0$ by dominated convergence. This concludes the proof. \qed
\bel \label{2Pttv} 
Let $e_{(0)}\df  (1,0)$, $e_{(1)}\df  (0,1)$, $e_{(\pm)}\df e_{(0)}\pm e_{(1)}$, be vectors in $\real^2$.  Furthermore, let $t\mapsto \rr_t\in \real^2$ be a family of vectors s.t. 
$\|\rr_t\|=O(|t|^{-\al})$ for some $0<\al<1$ in the Euclidean norm of $\real^2$. 
Then, for any fixed $f\in \mathcal{D}(\real^2)$ and $t\to \pm\infty$, there holds 
\begin{align} 
\int  \TwoPt_\reg(x-t(e_{(1)}+\rr_t)) f(x) d^2x& = 
\frac{-1}{4\pi} 2\ln(|t|\mu_v)q_f \, +\, O(|t|^{-\al} ),  \label{eq2Ptxte-one} \\
\int  \TwoPt_\reg(x-t(e_{(0)}+\rr_t  ) )  f(x) d^2x & = 
\frac{-1}{4\pi} 2(\ln(|t|\mu_v)\mp i\fr{\pi}{2})q_f  \, +\, O(|t|^{-\al}), \label{w-timelike-as} \\
\int \TwoPt_v(x-t e_{(\la)} ) f(x) d^2x &\! =\! 
\int \TwoPt_{v}^{-\lambda}(x)f(x) d^2x\! -\! \frac{1}{4\pi} \big\{\ln(2\mu_v |t|) \mp i \frac{\pi}{2}\big\}q_f+O(|t|^{-1}), \label{w-lightlike-as}
\end{align}
where $\TwoPt_{v}^\pm(x)  \doteq \lim_{\eps \downarrow 0} \frac{-1}{4\pi}\ln\big[i\mu_v x^\pm+\eps\big].$ 
\eel
\proof We will apply Lemma~\ref{log-lemma} considering that $\TwoPt_{\reg}(x)= \lim_{\eps\downarrow 0}\frac{-1}{4\pi} \ln(-\mu_v^2 x^2 + i \eps  x^0)$. In the case of~(\ref{eq2Ptxte-one}), we have $\TwoPt_\reg(x-t(e_{(1)}+\rr_t))= \lim_{\eps\downarrow 0} \frac{-1}{4\pi}\ln[h_1(x)+i\eps h_2(x)] $, where $h_1(x)=-\mu_v^2(x-t(e_{(1)}+\rr_t))^2$ and
$h_2(x)=(x-t(e_{(1)}+\rr_t))^0$. Thus, for any $f\in \mathcal{D}(\real^2)$,
\beqa
\int \TwoPt_\reg(x-t(e_{(1)}+\rr_t)) f(x)d^2x=\frac{-1}{4\pi} \int \big(  \ln|h_1(x)|  + \theta(-h_1(x)) \mrm{sgn}(h_2(x)) i\pi \big)f(x) d^2x.   \label{two-point-formula}
\eeqa
As the following equality holds
\beqa
h_1(x)=-\mu_v^2(x-t(e_{(1)}+\rr_t))^2= t^2\mu_v^2\bigg(1+
\underbrace{\frac{2(x-t\rr_t)\cdot e_{(1)}}{t}-\frac{(x-t\rr_t )^2}{t^2}}
\bigg),\label{h-1-formula}
\eeqa
for any compactly supported smearing function $f$ we can choose $t$ large enough so that the underbraced term has modulus smaller than $\frac{1}{2}$ and in particular $h_1(x)> 0$ on the support of $f$.
Thus we can omit the term involving $\theta(-h_1(x))$ in (\ref{two-point-formula}). We are left with
\beqa
\int \TwoPt_\reg(x-t(e_{(1)}+\rr_t)) f(x) d^2x= \frac{-1}{4\pi}  \ln[t^2\mu_v^2]q_f+O(t^{-\al}),
\eeqa
where we used $|\ln (1+y)| \leq (2 \ln 2) |y|$ for $|y|\leq \frac{1}{2}$.


In the case of (\ref{w-timelike-as}), we have $\TwoPt_\reg(x-t(e_{(0)}+\rr_t))= \lim_{\eps\downarrow 0} \frac{-1}{4\pi}\ln[h_1(x)+i\eps h_2(x)] $, 
where $h_1(x)=-\mu_v^2(x-t(e_{(0)}+\rr_t))^2$, $h_2(x)=(x-t(e_{(0)}+\rr_t))^0$. In this case we have
\beqa
h_1(x)= -t^2\mu_v^2\bigg(1-\frac{2(x-t\rr_t)\cdot e_{(0)}}{t}+\frac{(x-t\rr_t )^2}{t^2} \bigg).
\eeqa
By choosing $t$ sufficiently large, we obtain  that $h_1(x)< 0$ and $\mrm{sgn}(h_2(x))=\mrm{sgn}(-t)$ for all $x\in \supp(f)$. Thus a counterpart of
formula (\ref{two-point-formula}) gives
\beqa
\int \TwoPt_\reg(x-t(e_{(0)}+\rr_t)) f(x) d^2x= \frac{-1}{4\pi}  \big(\ln[t^2\mu_v^2]+\mrm{sgn}(-t)i\pi\big)  +O(t^{-\al}).
\eeqa

As for (\ref{w-lightlike-as}), we proceed as follows:
Write $x^{\pm} \doteq x^0 \pm x^1$ and note
\beqa
(x-t e_{(\lambda)})^2=x\cdot x- 2tx^{-\lambda} = -2tx^{-\lambda}\big(1-\frac{x^{\lambda}}{2t}\big),
\eeqa
where we have used that $x\cdot te_{(\lambda)}=t x^{-\lambda}$ and $x\cdot x = x^{\lambda}x^{-\lambda}$.
Then by equation~\eqref{eq2PtReg},
for $|t|$ sufficiently large, depending on the support of the smearing function $f$  (which we omit below in the notation, i.e., the limit is in $\mathcal{D}'(\real^2)$)
\begin{align} 
\TwoPt_\reg(x-te_{(\lambda)})  & = 
\lim_{\eps\downarrow 0}\frac{-1}{4\pi} \ln\big[ -\mu_v^2 (x-te_{(\lambda)})^2 + i\eps\,(x-te_{(\lambda)})^0\big] \non\\
&= \lim_{\eps\downarrow 0} \frac{-1}{4\pi} \ln\big[  2 t\mu_v^2 x^{-\lambda}\big(1-\frac{x^{\lambda}}{2t}\big)+ i\eps (x-te_{(\lambda)})^0  \big] \nonumber \\
&=\lim_{\eps\downarrow 0}\frac{-1}{4\pi} \bigg(\ln\big|h_1(x)\big|- \theta(-h_1(x))\mrm{sgn}(t )i\pi  \bigg)\non\\
&=\lim_{\eps\downarrow 0}\frac{-1}{4\pi} \bigg(\! \ln\big|2 t\mu_v \big|\!+\!\ln|\mu_v x^{-\lambda}|\!+\!\ln\big|1-\frac{x^{\lambda}}{2t}  \big| - \theta(-t\mu_v x^{-\la} )\mrm{sgn}(t )i\pi  \!\bigg)
\end{align}
for $h_1(x)= 2 t\mu_v^2 x^{-\lambda}\big(1-\frac{x^{\lambda}}{2t}\big)$ and $h_2(x)=(x-te_{(\lambda)})^0$ with $\mrm{sgn}(h_2)= -\mrm{sgn}(t)$.

On the other hand, we observe that 
 for $u\in\real$ 
\beq
\ln|u|- \theta(-tu) \sgn(t) i\pi = \lim_{\eps\downarrow 0} \ln(iu+\eps) - \sgn(t) i\frac{\pi}{2}.
\eeq
For this is equivalent to the  relation 
\beqa
\lim_{\eps\downarrow 0} \ln(iu+\eps)=\ln|u|+i\frac{\pi}{2}\big(\sgn(t)(1-2\theta(-tu) \big)=
\ln|u|+i\frac{\pi}{2}\sgn(u)\,. 
\eeqa
(We used the fact that $\sgn(t)(1-2\theta(-tu))=\sgn(u)$.)   
Thus we can write
\begin{align}
  \int \TwoPt_\reg(x-te_{(\lambda)}) f(x) d^2x & 
= \lim_{\eps\downarrow 0}\frac{-1}{4\pi}\int  \big(\ln\big|2 t\mu_v \big|+\ln[i\mu_v x^{-\lambda}+\eps]-   \sgn(t) i\fr{\pi}{2}\big) f(x)d^2x +O(t^{-1}) .
\end{align}
This concludes the proof.  \qed

\section{Computation of the scattering amplitude}\label{last-section}
\setcounter{equation}{0}

In this section we prove Theorem~\ref{main-result}. We choose $F=\pi(W(f))$, $G=\pi(W(g))$  as in the statement of Theorem~\ref{main-result} and 
denote by $S_{T}(F,G)$ the approximating sequence of the scattering amplitude from~(\ref{amplitude-def}). 
To simplify the expressions, we write
\beqa
f^{(1)}\df -g, \quad f^{(2)} \df -f, \quad f^{(3)}\df  f, \quad f^{(4)}\df g.
\eeqa
The corresponding charges are
\beqa
q_1\df -q_g, \quad q_2\df -q_f, \quad q_3\df q_f,\quad q_4\df  q_g.
\eeqa
We have by the Wick theorem (\ref{eqWickThmRegVEV}):
\beqa
& &S_T(F,G)\non\\
\!& =&\1\fr{   \int d^4t\, h(t_1)\ldots h(t_4)   \lan  :W(f^{(1)}_{(t^T_1)_-}):_v :W(f^{(2)}_{(t^T_2)_+} ):_v :W(f^{(3)}_{-(t^T_3)_+}) :_v :W(f^{(4)}_{-(t^T_4)_-}):_v\ran } 
{  \int d^4\tau\, h(\tau_1)\ldots h(\tau_4)  \lan  :W(f^{(1)}_{(\tau^T_1)_-}):_v :W(f^{(4)}_{-(\tau^T_4)_-}):_v \ran   \lan  :W(f^{(2)}_{(\tau^T_2)_+} ):_v :W(f^{(3)}_{-(\tau^T_3)_+}) :_v    \ran} \non\\
\1&=&\1\fr{ \phantom{44}\int d^4t\, h(t_1)\ldots h(t_4) \begin{Bmatrix} \lan  :W(f^{(1)}_{(t^T_1)_-}):_v :W(f^{(2)}_{(t^T_2)_+}  ):_v\ran_v  
 \lan :W( f^{(1)}_{(t^T_1)_-} ):_v  :W( f^{(3)}_{-(t^T_3)_+} ) :_v\ran_v \\
\times  {\gc \lan  :W(f^{(1)}_{(t^T_1)_-}):_v :W(f^{(4)}_{-(t^T_4)_-}):_v \ran_v   \lan  :W(f^{(2)}_{(t^T_2)_+} ):_v :W(f^{(3)}_{-(t^T_3)_+}) :_v    \ran_v} \\
 \times \lan :W(f^{(2)}_{(t^T_2)_+} ):_v :W( f^{(4)}_{-(t^T_4)_-}  ):_v \ran_v     \lan  :W( f^{(3)}_{-(t^T_3)_+}  ) :_v :W(f^{(4)}_{-(t^T_4)_-} ):_v\ran_v  \end{Bmatrix}  }
{\int d^4\tau\, h(\tau_1)\ldots h(\tau_4) {\gc \lan  :W(f^{(1)}_{(\tau^T_1)_-}):_v :W(f^{(4)}_{-(\tau^T_4)_-}):_v \ran_v   \lan  :W(f^{(2)}_{(\tau^T_2)_+} ):_v :W(f^{(3)}_{-(\tau^T_3)_+}) :_v    \ran_v}},  \quad\quad \label{long-formula}
\eeqa
where $t^T_i=T+s(T)t_i$.   (Actually, for negative times we  get $(t^T_i)=T-s(T)t_i$, but this can be readjusted  
using that $h$ is symmetric). We note that the middle line in the numerator of (\ref{long-formula}) has the same structure 
as the denominator and only the time averaging prevents immediate cancellation. As we will see, the cancellation actually takes
place in the limit $T\to \infty$.
Let us now list the differences of time arguments appearing in the numerator of formula~(\ref{long-formula}):
\beqa 
& &(t_1^T)_- -(t^T_2)_+=2T\bigg( (0,-1)-\h \fr{s(T)}{T} \big( t_1-t_2, t_1+t_2 \big)   \bigg)=-2T(e_{(1)}+\rr_T^{t_1,t_2}),\\ 
& & (t^T_1)_- +(t^T_3)_+=2T\bigg(  (1,0)+ \h \fr{s(T)}{T}(t_1+t_3, t_3-t_1)  \bigg) =2T(e_{(0)}+\rr_T^{t_1,t_3}), \\
& & {\gc  (t_1^T)_-+(t_4^T)_-=2T\bigg( (1,-1)+\fr{s(T)}{2T}  \big(  t_1+t_4, -(t_1+t_4)   \big)  \bigg) =2T\bigg(1 +\fr{s(T)}{2T} (t_1+t_4)   \bigg)e_{(-)}  }, \quad\quad \\ 
& & {\gc  (t^T_2)_+ +(t^T_3)_+=2T\bigg( (1,1)+\fr{s(T)}{2T}\big(  t_2+t_3, t_2+t_3   \big) \bigg)= 2T\bigg(1 +\fr{s(T)}{2T} (t_2+t_3)   \bigg)e_{(+)} }, \\ 
& &(t_2^T)_++(t_4^T)_-=2T \bigg( (1,0)+\h \fr{s(T)}{T} \big( t_4+t_2, t_2-t_4 \big)   \bigg)=2T(e_{(0)}+\rr_T^{t_2,t_4}),\\
& &-(t_3^T)_++(t_4^T)_-=2T \bigg( (0,-1)-\h \fr{s(T)}{T} \big( t_4-t_3, t_4+t_3 \big)   \bigg)=-2T(e_{(1)}+\rr_T^{t_3,t_4}).  
\eeqa
Here the vectors $r_T^{t_i,t_j}$  differ from line to line, but they satisfy $\| r_T^{t_i,t_j}\|=O(T^{-\al})$, $0<\al<1$, uniformly 
on compact sets in $t_i, t_j$. 

Concerning the first line in the numerator of formula~(\ref{long-formula}), we obtain from Lemma~\ref{2Pttv}
\beqa
& & \lan  :W(f^{(1)}_{(t^T_1)_-}):_v :W(f^{(2)}_{ (t^T_2)_+ } ):_v\ran_v = \exp\big(- w_v(f^{(1)}_{(t^T_1)_- }  ,f^{(2)}_{(t^T_2)_+}) \big)\non\\
& &\phantom{4444} = \exp\big(-\int f^{(1)}(x)w_v\big(x-y+((t^T_1)_-- (t^T_2)_+)\big) f^{(2)}(y)\, d^2xd^2y  \big) \non\\
& &\phantom{4444}= \exp\big(  \frac{1}{4\pi} 2\ln( 2T\mu_v)\underbrace{q_1q_2}_{q_fq_g} \, +\, O(T^{-\al}) \big), \label{one}  \\
& & \lan :W( f^{(1)}_{(t^T_1)_-} ):_v  :W( f^{(3)}_{-(t^T_3)_+} ) :_v\ran_v = \exp\big(- w_v(f^{(1)}_{(t^T_1)_- }  ,f^{(3)}_{-(t^T_3)_+}) \big)\non\\
& &\phantom{4444} = \exp\big(-\int f^{(1)}(x)w_v\big(x-y+((t^T_1)_-+ (t^T_3)_+)\big) f^{(3)}(y)\, d^2xd^2y  \big)  \non\\
& &\phantom{4444}= \exp\big(  \frac{1}{4\pi} 2 \big(\ln( 2T\mu_v) +i\fr{\pi}{2}  \big)\underbrace{q_1q_3}_{-q_fq_g} \, +\, O(T^{-\al}) \big). \label{x}
\eeqa
In the  the middle line in the numerator of formula~(\ref{long-formula}), as well as in its denominator, we have translations in lightlike directions. We set  
$\be_T^{t_i,t_j}\df 1 +\fr{s(T)}{2T} (t_i+t_j)$  and compute using Lemma~\ref{2Pttv}
\beqa
& &\lan  :W(f^{(1)}_{(t^T_1)_-}):_v :W(f^{(4)}_{-(t^T_4)_-}):_v \ran_v =\exp\big(- w_v(f^{(1)}_{(t^T_1)_- }  , f^{(4)}_{-(t^T_4)_-}   ) \big) \non\\
& & \phantom{4444} = \exp\big(-\int f^{(1)}(x)w_v\big(x-y+((t^T_1)_-+ (t^T_4)_-)\big) f^{(4)}(y)\, d^2xd^2y  \big) \non\\
& &\phantom{4444}= \exp(-w_{v}^+(f^{(1)}, f^{(4)}) +\fr{1}{4\pi}\{  \ln(4\mu_v T \be_T^{t_1,t_4} ) +i    \fr{\pi}{2}  \} \underbrace{q_1q_4}_{-q_g^2 } +O(T^{-1})  ), \\
& & \lan  :W(f^{(2)}_{(t^T_2)_+} ):_v :W(f^{(3)}_{-(t^T_3)_+}) :_v    \ran_v= \exp\big(- w_v(f^{(2)}_{(t^T_2)_+}  , f^{(3)}_{-(t^T_3)_+}   ) \big) \non\\
& & \phantom{4444} = \exp\big(-\int f^{(2)}(x)w_v\big(x-y+((t^T_2)_++ (t^T_3)_+)\big) f^{(3)}(y)\, d^2xd^2y  \big) \non\\
& &\phantom{4444}= \exp(-w_{v}^-(f^{(2)}, f^{(3)}) +\fr{1}{4\pi}\{  \ln(4\mu_v T \be_T^{t_2,t_3} ) +i    \fr{\pi}{2}  \} \underbrace{q_2q_3}_{-q_f^2} +O(T^{-1})  ). 
\eeqa 
Finally, in the bottom line of the numerator of formula~(\ref{long-formula}) we have
\beqa
& & \lan  :W(f^{(2)}_{(t^T_2)_+}):_v :W(f^{(4)}_{-(t^T_4)_-} ):_v\ran_v = \exp\big(- w_v(f^{(2)}_{ (t^T_2)_+},f^{(4)}_{- (t^T_4)_-}) \big)\non\\
& &\phantom{4444} = \exp\big( -\int f^{(2)}(x)w_v\big(x-y+((t^T_2)_+ +(t^T_4)_-)\big) f^{(4)}(y)\, d^2xd^2y  \big)\non\\
& &\phantom{4444}= \exp\big(  \frac{1}{4\pi} 2\big(\ln(2T\mu_v)+ i\fr{\pi}{2}\big) \underbrace{q_2q_4}_{-q_gq_f} \, +\, O(T^{-\al})  \big), \label{four} \\
& & \lan  :W(f^{(3)}_{-(t_3^T)_+ }) :_v :W(f^{(4)}_{-(t^T_4)_-}):_v\ran_v = \exp\big(-w_v(f^{(3)} _{-(t_3^T)_+}, f^{(4)}_{ -(t^T_4)_- } ) \big) \non\\
& &\phantom{4444} =\exp\big( -\int f^{(3)}(x) w_v(x-y+( -(t_3^T)_++ (t^T_4)_-) ) f^{(4)}(y )  \, d^2xd^2y  \big)\non\\
& &\phantom{4444}=\exp\big(  \frac{1}{4\pi} 2\ln( 2T\mu_v)\underbrace{q_3q_4}_{q_gq_f} \, +\, O(T^{-\al}) \big). \label{two}
\eeqa
Now substituting (\ref{one})--(\ref{two}) to (\ref{long-formula}) we note the cancellation of all terms involving $ \frac{1}{4\pi} 2\ln( 2T\mu_v)q_gq_f$ in the numerator
and the contributions $\exp( - \fr{1}{4\pi}\{  \ln(4\mu_v T) +i    \fr{\pi}{2} \}) q_{f/g}^2$ between the numerator and the denominator.
Thus we obtain
\beqa
S_T(F,G)
= e^{-\fr{i}{2}q_fq_g } \fr{  \int d^4t\, h(t_1)\ldots h(t_4)  
\exp\big(- \fr{ q_g^2 }{4\pi}  \ln(\be_T^{t_1,t_4} )  - \fr{ q_f^2}{4\pi}  \ln(\be_T^{t_2,t_3} )      + O(T^{-\al})  \big)   } 
{  \int d^4\tau\, h(\tau_1)\ldots h(\tau_4)  \exp\big(  - \fr{ q_g^2 }{4\pi}  \ln(\be_T^{\tau_1,\tau_4} )  - \fr{ q_f^2}{4\pi}  \ln(\be_T^{\tau_2,\tau_3} )   + O(T^{-\al})  \big)},
\eeqa
where the crucial pre-factor $e^{-\fr{i}{2}q_fq_g }$ originates from the $i\fr{\pi}{2}$-terms in (\ref{x}), (\ref{four}). 

Now since $h$ is compactly supported and $\lim_{T\to\infty}\be_T^{t,t'}= \lim_{T\to\infty}  \big( 1 +\fr{s(T)}{2T} (t+t')\big)= 1$ for any $t,t'\in\real$, the dominated convergence gives
\beqa
\lim_{T\to\infty}S_T(F,G)= e^{-\fr{i}{2}q_fq_g}.
\eeqa
This concludes the proof. \qed

\appendix
\renewcommand{\thesubsection}{\thesection.\arabic{subsection}}
\setcounter{equation}{0}
\renewcommand{\theequation}{\thesection.\arabic{equation}}
\setcounter{Thm}{0}
\renewcommand{\theThm}{\thesection\,\arabic{Thm}}

\section{Spectral properties of the vacuum}\label{GNS}

Continuity of the unitary group implementing the Poincar\'e  transformations follows 
from the continuity of the following function in some neighbourhood of unity in the Poincar\'e group
\beqa
 L\mapsto \lan  \Om, \pi(W(f)) U(L)\pi(W(g))\Om \ran\!=\!\lan W(f) W(g_L)\ran\!=\!\de_{q_f+q_g,0} \lan W(f)\ran_v \lan W(g)\ran_v
  e^{- w_v(f, g_{L}) }.  \label{Poincare-action} \,\,
  \eeqa
Here $f,g\in \mathcal{D}_{\real}(\real^2)$ and we made use of (\ref{eqWickThmRegVEV}).
The continuity of $L\mapsto w_v(f, g_{L})$ can be conveniently checked using formulas~(\ref{eq2PtReg}), (\ref{ln-lemma})
and the dominated convergence. Now  restricting attention to translations in (\ref{Poincare-action}), one can determine 
the support of the Fourier transform of this function using its analyticity properties via \cite[Theorem~IX.16]{RSII}.
This gives the spectrum condition, i.e., the spectral measure of  $a\mapsto U(a)$ is supported in the future lightcone.

Proceeding to more detailed properties of the spectrum, let us show the inclusion (\ref{waves-are-neutral}), i.e., neutrality of the
waves.  By the ergodic theorem, the orthogonal projection on the single-wave subspaces $\hil_{\la}$, $\la=\pm$, can be written as follows:
\beqa
E_{\la}=\slim_{T\to \infty} \int dt\, h_T(t) U(t, \la t). 
\eeqa 
Thus to prove (\ref{waves-are-neutral}), it suffices to show that  
\beqa
\lim_{t\to \infty} \lan \Om, \pi(W(g))\pi(W(f_{(t,\la t)} ))\Om\ran=0 \label{scalar-product}
\eeqa
unless $q_{g}=q_f=0$. Suppose that one of the charges is different from zero.
Then the scalar product in (\ref{scalar-product}) is zero unless $q_g+g_f=0$.
Now making use of the lightlike asymptotics in (\ref{w-lightlike-as}), we can write
\beqa
& &\lan \Om, \pi(W(g)\pi(W(f_{(t,\la t)} ))\Om\ran\non\\
& &=\lan W(g)\ran_{v} \lan W(f)\ran_{v} 
e^{-\big( \int \TwoPt_{v}^{-\lambda}(x)(g*f_-)(x) d^2x - \frac{1}{4\pi} \big\{\ln(2\mu_v|t|) - i \frac{\pi}{2}\big\}q_{f}q_g +O(|t|^{-1}) \big) },
\eeqa
where $(g*f_-)(x)\df \int g(x-y)f(-y)d^2y$. Since $q_{f}q_g<0$, the above expression tends to zero with $t\to\infty$.

\section{Regularized two-point function}  \label{2PtBoson}
\newcommand{\kk}{p} 

\bel Under the assumptions on $v$ specified in Section~\ref{Asymptotics}
\begin{align} \label{eq2PtReg''}
\TwoPt_{\reg}(x) & \doteq  (2\pi)^{-1}\int d\mu_0(p) \big(e^{-ip\cdot x} - v(p)\big) 
= \lim_{\eps\downarrow 0}\frac{-1}{4\pi} \ln(-\mu_v^2 x^2 + i \eps x^0), \\
\TwoPt_{v}^\pm(x) & \doteq  (2\pi)^{-1} \int_{{ \kk^1} { \lessgtr }   0} d\mu_{0}(\kk)\big(e^{-i\kk\cdot x} - v(\kk) \big)
= \lim_{\eps\downarrow 0}\frac{-1}{4\pi} \ln(i \mu_v x^\pm +  \eps) \,.\label{eq2PtChir} 
\end{align}
\eel
\proof We consider first 
${ {\kk^1}>0}$ in \eqref{eq2PtChir}. In this case, $\kk\cdot x = |\kk^1|x^0{ -}\kk^1x^1= \kk^1 x^{ -}$,  where $x^{ -}=x^0 -x^1$. Thus, the LHS of \eqref{eq2PtChir} is $-(4\pi)^{-1} I( { x^-})$, with 
\beq \label{eqI}
I(u) = \int_0^\infty \frac{d\kk^1}{\kk^1}\big(v({ \kk^1}, { \kk^1})-e^{-i\kk^1 u} \big).
\eeq
(This is meant as a distribution on $\RR$, i.e., integrate up to ${ \kk^1}=K$,  smear in u with a test function and then take the limit $K\to \infty$).  
First, note that  by a simple computation, the scaling degree of $I$ is zero. 
Thus, it is uniquely fixed by its restriction to $\RR\setminus \{0\}$, cf. \cite[Theorem 7.2]{FR13}.   
Its restriction $I_+$ to $\calD(\RR^+)$ has the imaginary part
$\Im I_+(u) =  \int_0^{\infty} \fr{d{ \kk^1}}{{ \kk^1}} \sin({ \kk^1})=\frac{\pi}{2}$ 
and satisfies the ODE $(I_+)'(u) = 1/u$. 
This implies that 
\beqa
I_+(u) = \ln(\mu_+ u) + i\frac{\pi}{2} = \ln(i \mu_+ u) 
\eeqa
for some constant $\mu_+>0$.  Here we used that also for distributions $T'=0$ implies $T=\mrm{const}$ \cite[p. 51]{Schwartz}. The real part of the constant is $\ln(\mu_+)$, whereas 
the imaginary part is $\fr{\pi}{2}$.
Similarly, one finds that the restriction $I_-$ to $\calD(\RR_-)$ is $I_-(u)=\ln(i\mu_- u)$ for some $\mu_-$. But the relation  $I_-(-u)=\overline{I_+(u)}$, that follows from \eqref{eqI}, implies that $\mu_+=\mu_-\doteq \mu_v$. 
We conclude that $I(u)=\ln(\mu_v |u|) +\mrm{sgn}(u) i\frac{\pi}{2}={ \lim_{\eps\downarrow 0} } \ln(i\mu_v u+\eps)$ on $\RR$. This proves equation~\eqref{eq2PtChir} for the case ${ \kk^1>0}$.

In the case ${ \kk^1<0}$, $\kk\cdot x = { -\kk^1x^0-\kk^1x^1= -\kk^1 x^+}$, and the LHS of \eqref{eq2PtChir} is
\beqa
\fr{-1}{4\pi} \int_{-\infty}^0 \frac{{ d\kk^1}}{|\kk^1|}\big(v(|\kk^1|,\kk^1)-e^{i\kk^1 x^{ +}} \big)
= \fr{-1}{4\pi}\int_0^\infty \frac{d\kk^1}{\kk^1}\big(v(\kk^1,-\kk^1)-e^{-i\kk^1 x^{ +}} \big)=\fr{-1}{4\pi} I(x^{  +}),
\eeqa
since we assumed $v(|\kk^1|,\kk^1)=v(|\kk^1|,-\kk^1)$.

Equation~\eqref{eq2PtReg''} follows from $w_v(x)=w_{v}^{+}(x) +  w_{v}^{-}(x)$ .  More precisely, we have 
\beqa
w_v(x)=w_{v}^{+}(x) +  w_{v}^{-}(x) \1&=&\1  \fr{-1}{4\pi} \bigg(  \ln(\mu_v| x^-|) +\ln(\mu_v|x^+| )+\h (\mrm{sgn}(x^-)+ \mrm{sgn}(x^+)) i\pi \bigg)\non\\
\1&=&\1 \fr{-1}{4\pi} \bigg(  \ln(\mu_v^2 | x^2|) +\theta(x^2) \mrm{sgn}(x^0) i\pi \bigg) \non\\ 
\1&=&\1  \lim_{\eps\downarrow 0} \frac{-1}{4\pi}  \ln(-\mu_v^2 x^2 + i \eps x^0),  
\eeqa
where in the second step one simply checks all the possibilities and in the last step we used Lemma~\ref{log-lemma}. \qed

\end{document}